\documentclass[usenatbib, usegraphicx]{mn2e} 

\title[An Iterative Method for Constructing Equilibrium Phase Models 
of Stellar Systems]
{An Iterative Method for Constructing Equilibrium Phase Models of Stellar
Systems} 

\author[S.A.~Rodionov, E.~Athanassoula, N.Ya~Sotnikova]
{S.A.~Rodionov$^{1}$\thanks{E-mail: seger@astro.spbu.ru}, 
E.~Athanassoula$^{2}$
and N.Ya.~Sotnikova$^{1}$\\
$^{1}$Sobolev Astronomical Institute, 
St. Petersburg State University, 
Universitetskij pr.~28, 198504 St. Petersburg, Stary Peterhof, Russia\\
$^{2}$ 
Laboratoire d'Astrophysique de Marseille (LAM), UMR6110,
CNRS/Universit\'e de Provence, Technop\^ole de Marseille-Etoile, \\
38 rue Fr\'ed\'eric Joliot Curie, 13388 Marseille C\'edex 20, France} 

\date{Accepted ???? ??? ??. Received ???? ??? ??; in original form ???? ??? ??}

\pagerange{\pageref{firstpage}--\pageref{lastpage}} \pubyear{2008}

\begin{document}
\label{firstpage}

\maketitle

\begin{abstract}
We present a new method for constructing equilibrium phase models for
stellar systems, which we call the iterative method. It relies on
constrained, or guided evolution, so that the equilibrium solution has
a number of desired parameters and/or constraints. This method is very
powerful, to a large extent due to its simplicity. It can be used for
mass distributions with an arbitrary geometry and a large variety of
kinematical constraints. We present several examples illustrating it. 
Applications of this method include the creation of initial conditions
for $N$-body simulations and the modelling of galaxies from their
photometric and kinematic observations. 
\end{abstract}


%----------------------------------------------------------------------------%
\section{Introduction}

In astronomy there are at least two problems where  
equilibrium phase models of stellar systems need to be constructed. 
The first one is the construction of phase models 
for real galaxies from observational data, i.e. the modelling of
observational data. The second problem is the construction of initial
conditions for $N$-body simulations of stellar systems. It is obvious
that these two problems are tightly connected, and yet they have, so
far, been solved by different methods. The Schwarzschild method \citep{S79} 
and its modifications is often used for
modelling of observational data \citep[e.g][]{H00, vdB06, T07, vdB08, deL08},
but has almost never been used so far 
to produce initial conditions for simulations. For $N$-body initial
conditions, a wide variety of methods has been used, based on the
Jeans theorem \citep[e.g. ][]{Zangthesis, AS86, KD95, WD05, M07}, or on Jeans'
equations \citep[e.g. ][]{H93}. In the case of multi-component systems,
e.g. disc galaxies with a bulge and a halo, the components are built
separately and then either simply superposed, or the potential of the
one is adiabatically grown in the other \citep[e.g. ][]{Barnes88, M07, A07}
before superposition.

For real galaxies the phase space density is generally unknown, but
we do have some information about it. For example, we know more or
less accurately a distribution of mass for the visible components
(notwithstanding uncertainties due to the mass to light ratio) and we
often have some constraints on the velocity distribution. It is also
reasonable to assume that the galaxy is in an 
equilibrium state. So in general, the problem of constructing a
model in phase space is equivalent to constructing an equilibrium
phase model with a given mass distribution and, in many cases, given
kinematic constraints. In the case of modelling observational data
(first of the two above mentioned problems) the kinematic parameters are
the observed velocities integrated along the line of sight. In the case of
initial conditions 
for $N$-body simulations, a wide variety of kinematic parameters is
possible, depending on the problem the simulation addresses.

We have developed a new method for constructing equilibrium phase
models with a given mass distribution and with given kinematic parameters, 
which we call the iterative method. It can be applied to systems with
arbitrary geometry, so that the requested mass distribution can be  
arbitrary. The idea and a first implementation was 
presented in \citet{RS06}. In \citet{RO08} we improved it, and applied
it to construct an $N$-body model of the stellar disk of our Galaxy for two
realistic mass models of the Milky Way. Here we present a final version of
this method, fully allowing kinematical constraints. In the previous
articles we had concentrated on constructing equilibrium phase models
with a given mass distribution, so that kinematic parameters were
either not considered or only in terms of auxiliary parameters, such
as the total angular momentum \citep{RS06, RO08}. This, however,
limited the applicability of our method, both for initial conditions
and for modelling real galaxies. Indeed, initial kinematics play a
crucial role in determining the evolution of $N$-body systems, while
observational constraints more often than not include kinematics.
In this paper we give equal attention to the mass distribution and
kinematical constraints, so that the iterative method can now be used
for a number of interesting applications. In principle, in our method,
both the kinematic constraints and the mass distribution can be
arbitrary. But the part of our algorithm that concerns the kinematic
constraints is not universal, contrary to the part that handles the
mass distribution, but is tailored to the specific constraint. 
Here we consider several types of constraints. 
Once these are understood, it is rather easy to
extend the algorithm for every new type of kinematic parameter (see
below).  

The power of the iterative method stems from its simplicity. The
iterative method is based on a simple and, in a way, obvious idea,
which is
implemented in a simple algorithm. The purpose of this article is to
fully describe this method. We first introduce the basic concept in
Section~\ref{s_imethod}, where we also explain the different modules of
the algorithm and the way they should be applied. In
section~\ref{s_models} we illustrate the use of the method with 
three
examples, namely a triaxial system, a multi-component model of a
disk galaxy (including live disk, bulge and halo components) and a disk
constructed with given line-of-sight kinematic. 
We briefly conclude in section~\ref{s_conc}.

\section{The Iterative Method}
\label{s_imethod}

\subsection{General Idea of the Iterative Method}
\label{generalidea}

The aim of the iterative method is to construct equilibrium $N$-body models
with a given mass distribution and with given kinematic properties, 
parameters, or constraints. This method relies on the fact that any
non-equilibrium system will tend, more or less fast, to a stable
equilibrium. We thus start by constructing any arbitrary, 
non-equilibrium $N$-body system, and let it evolve.  
Such an evolution changes both its mass distribution and its
kinematics, so that the final system does not have the desired properties.
To achieve the latter, we developed a new method which we call the
iterative method and which relies on a constrained, or guided,
evolution. We will describe it fully in this section. 
This idea is of course applicable for any arbitrary dynamical system
and is even widely used in every day life. 

To give an example, let us consider a donkey walking by itself in a
field. After some time the donkey can be anywhere in the field.
Now consider another donkey which we attach to a tree by a rope. This
donkey also walks in the field, but it will have to stay inside a circle of
radius equal to the length of the rope. This is an example of a
constrained evolution, which will necessarily lead to a final state
within a circle around the tree. The crucial point now is how to
achieve this constraint, i.e. what will the equivalent of the rope be
in the case of galaxy models.

\begin{figure}
\begin{center}
\includegraphics[width=8cm]{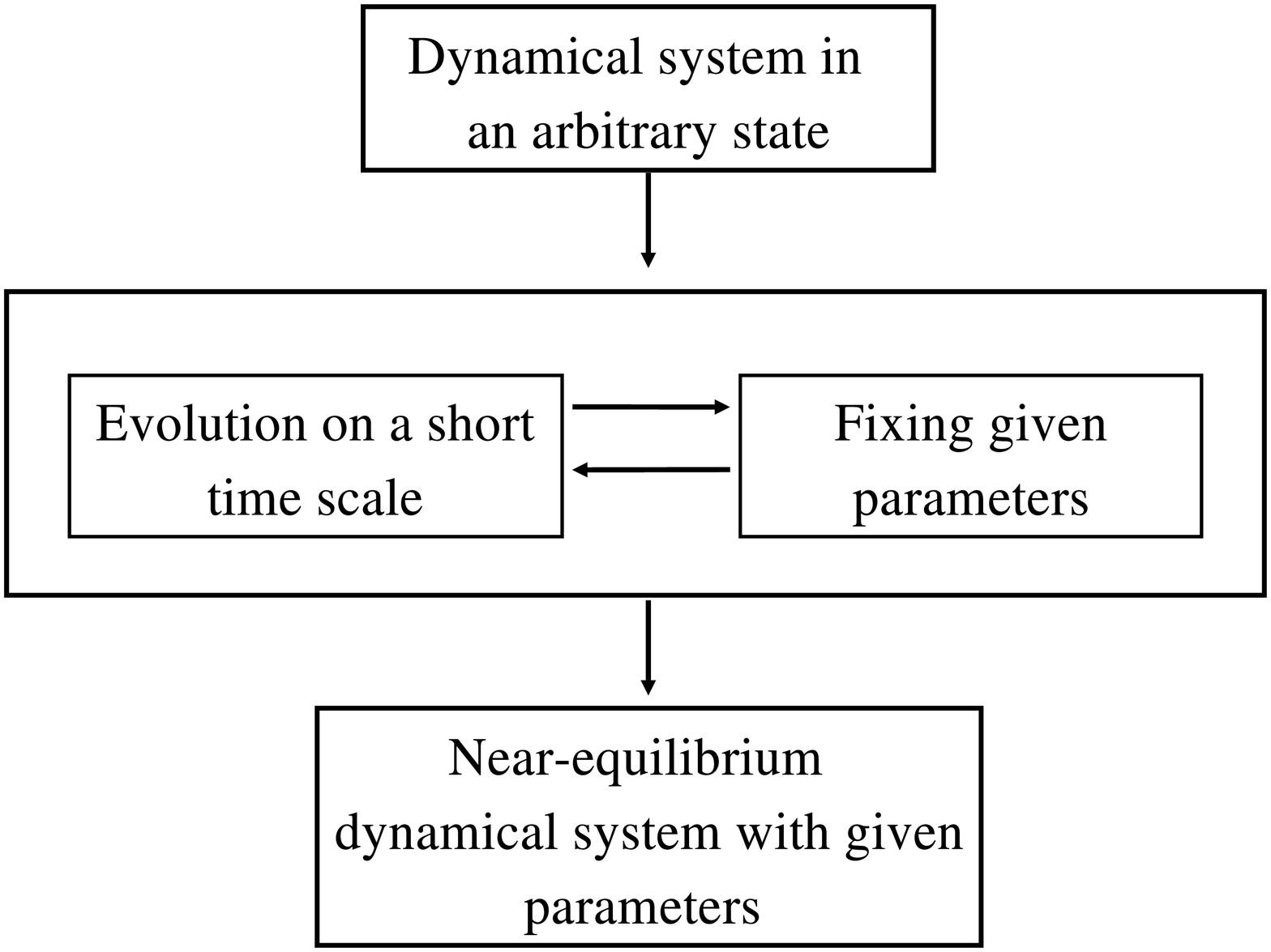}
\end{center}
\caption{General scheme of the iterative method in the case of an arbitrary
dynamical system.}
\label{fig_genscheme}
\end{figure}

The general scheme of our method is presented
in Fig.~\ref{fig_genscheme} and can be applied to any arbitrary
dynamical system. Let our task be to find an equilibrium state of some
dynamical system obeying given constraints or having specific values of
some given parameters. We start from any arbitrary state of our
dynamical system 
and allow the system to evolve during a short time interval. We then
need to make sure that the given parameters have the required values. 
In order to achieve this we need to modify the system so that the 
given parameters have the necessary values, while making sure that the
other quantities or parameters are kept unchanged, so as to retain
memory of the evolution. As shown in Fig.~\ref{fig_genscheme}, we
iterate these two steps,
alternating short evolutions and modifications of the system to fix
the set parameters. We 
thus constrain the evolution in order for it to reach an equilibrium
state with the desired set of constraints. We stop when we consider we
are sufficiently near the desired equilibrium state of the system.

\begin{figure}
\begin{center}
\includegraphics[width=8cm]{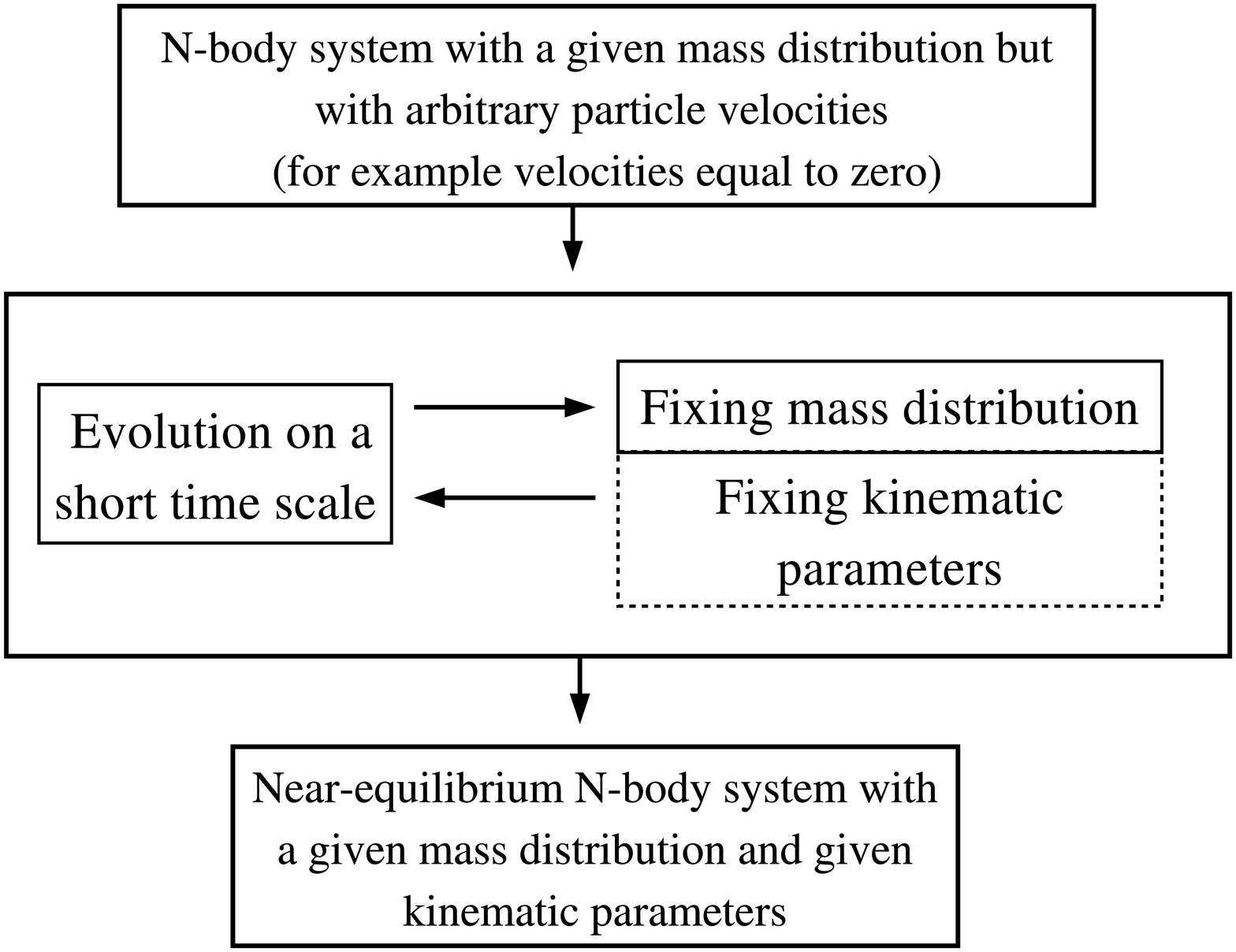}
\end{center}
\caption{The scheme of the iterative method for the case of an $N$-body
system with a given mass distribution and given kinematical
parameters.}
\label{fig_scheme}
\end{figure}

Let us now consider a case, in which we wish to construct an equilibrium
$N$-body system with a given mass distribution and with or without given
kinematic constraints. The scheme is outlined
schematically in Fig.~\ref{fig_scheme}. 
We initially create an $N$-body system with a given mass distribution but
with arbitrary particle velocities (for example velocities equal to zero).
 We then start 
the iterative procedure, by letting the system go through a sequence
of evolutionary steps of short duration. At the end of each one of
these steps, and before the new step is started, we need to set the
appropriate parameters. Let us first 
consider the case where we wish to have a specific mass distribution,
but have no kinematic constraints. To achieve this, we construct a new
$N$-body system, with the desired mass distribution but with
velocities chosen according to the velocity distribution obtained from
the evolution. In other words, we ``transfer'' 
the velocity distribution from the system obtained from the evolution
to a new system, which will have the desired mass
distribution and an evolved velocity distribution. 
The algorithm performing this ``transfer'' is the core of
the iterative method, and will be discussed in more
detail in section~\ref{s_transvel}. If we have kinematic constraints
as well, we need to modify the velocities of the particles in
such a way that the constraints are fulfilled and, at the same time, as
little as possible so that some memory of
the evolution is kept. How this is done in practise depends on the
imposed constraints and will be described, for a number of cases, 
in section~\ref{s_fixk}. Procedures for further types of
constraints can be easily found following similar techniques. In all
cases we have a new 
system which has the desired mass distribution, obeys the necessary
velocity constraints, while being nearer to equilibrium, since it
retains partial memory of the evolution. We repeat this iterational
procedure a number of times, alternating one evolution phase and one
phase where the necessary parameters are set, until we come as near to
the desired equilibrium state as desired.

\subsection{Transfer of velocity distribution}
\label{s_transvel}

The transfer problem can be formulated as follows. Any evolution step
ends with a model, which we will refer to as the ``old'' model. This
step is followed by a constraining, or fixing step, during which we
create a ``new'' model with the desired mass distribution. We now need
to transfer the velocity information from the ``old'' to the ``new''
model. There is more than one way to achieve this transfer. 
\citet{RS06} used an algorithm based on moments
of the velocity distribution, which, however, proved to
be rather complicated and cumbersome. Here we suggest a much simpler
and more reliable algorithm, which is in fact an improvement of an
algorithm initially used in \citet{RO08}. 

The basic idea of our velocity transfer algorithm is as follows.
We assign to the new-model particles the velocities of
those particles from the old model that are nearest to the ones in the new
model. The simplest (although, as we show below, not necessarily optimum)
implementation of this idea is evident. One can prescribe to each particle
in the new model the velocity of the nearest particle from the old model.
Let us formulate this proposition more strictly. For each $i$-th
particle of the new model, one finds the $j$-th particle in the old model 
with the minimum value 
of $|{\bf r}^{new}_i - {\bf r}^{old}_j|$. Here, ${\bf r}^{new}_i$ 
is the radius vector of the $i$-th particle
in the new model, and ${\bf r}^{old}_j$ is the radius vector of the 
$j$-th particle in the old model. Hereafter we will always imply this
definition when we talk of the nearest or closest particle. One then
takes as the velocity of the $i$-th particle in the new 
model the velocity of the $j$-th particle in the old model. This simple
algorithm has, however, one essential drawback. If the numbers of particles
in the old and new models are the same then only about one-half of the
particles in the old model participate in the velocity transfer. The reason
is that many old-model particles transfer their velocities to several
particles in the new model. As a result of this, almost one-half of 
the particles in the old model do not transfer their velocities at all. This
means that a significant amount of information on the velocity distribution
will be lost in the transfer process. The noise will therefore grow,
as we verified in numerical experiments. Thus, this transfer algorithm
is not optimum. 

This shortcoming can, nevertheless, be overcome by 
modifying this transfer scheme. 
For this, we introduce an input parameter, which we call
the ``number of neighbours'' $n_{nb}$. We also introduce, for each
particle in the old model, the parameter $n_{used}$, which denotes
the number of times this particle has been used for velocity copying. At the
beginning of the transfer procedure we set $n_{used}=0$ for each particle
in the old model, since its velocity information has not been yet
transferred to any of the new model particles. 
%We then calculate what the
%value of $n_{use}$ would be for each old model particle if the
%velocity assigned to this particle is simply that of its nearest
%neighbour in the new model, as is done in \citet{ROO08}. 
For any given particle in the new model we find the nearest $n_{nb}$
neighbours in the 
old model (the closeness being understood as defined above) and from
these we single out the subgroup of particles that have a minimum
$n_{used}$. From this subgroup we find the particle that is the closest 
one to the position of the new-model particle, add one to its
$n_{used}$ value and prescribe its velocity to the new-model particle
we are examining.

We note that if $n_{nb} = 1$ this algorithm is the same as the 
previous one and about half of the particles will not take part in
the velocity distribution transfer. If, however, we take
$n_{nb} = 10$, only a small fraction (a few per cent) of old-model
particles will not take part in the transfer process. We adopt this
improved transfer method since we showed that it gives good results 
in the iterative procedure.

If the desired model has some symmetry, it can be useful 
to take it into account in the algorithm of velocity transfer by
redefining the distance between two particles. For example, if we wish
to build an axisymmetric system, we 
search for the nearest particles in the
two-dimensional space $R-z$ (where $R$ the cylindrical radius) instead of
the three-dimensional space $x-y-z$. We then transfer the
velocity of this nearest old-model particle (in cylindrical
coordinates) to the new-model particle. It is important to
adopt this new definition of the distance in order to 
fix not only the mass distribution, but also fix the velocity
distribution and to make it fully axisymmetric.

We have thus introduced three variants of the velocity transfer
algorithm. We will refer to 
them as ``transvel\_3d'', ``transvel\_cyl'' and ``trasvel\_sph''.

\begin{enumerate}
\item 
``transvel\_3d'': This is the basic algorithm, for the case when the
  desired system has no assumed symmetry. By using this algorithm in the
iterative method we only fix the mass distribution and leave the velocity
distribution unchanged. In the current work we use this algorithm when 
constructing triaxial models.

\item
``trasnvel\_cyl'': This is a modification of the basic algorithm for
axisymmetric systems and was described just above. We use this
algorithm when both the desired mass and velocity distribution are 
axisymmetric. In the current
work we use this algorithm for constructing all models except for
the triaxial ones.  

\item
``transvel\_sph'': This is a modification of the basic algorithm for
  spherical systems. In this version of the algorithm we search for
  the nearest particles in one dimensional ``r'' space, where $r$ is
  the spherical radius. By using this algorithm in the
  iterative method we fix both the mass and the velocity distribution to 
  make the system fully spherically symmetric. This was used
  in \citet{SR08}. 

\end{enumerate}

\subsection{Fixing the Kinematic Parameters}
\label{s_fixk}
Here we describe algorithms for fixing different kinematic parameters.
The general algorithm is as follows. We slightly change the particle 
velocities to fix given kinematic parameters, but we keep as many other
parameters as possible unchanged. Here we describe in detail only 
a number of algorithms, which we use in this paper. But it is easy to
develop algorithms for any other kinematic parameter. It 
is only necessary to follow the general principle: ``keep unchanged
the parameters that do not need to be fixed''.

\subsubsection{Fixing the radial velocity dispersion profile $\sigma_R(R)$ }
\label{s_fixsvR_R}

We use this algorithm in order to fix in stellar disks the radial
velocity dispersion profile to a given function $\sigma_R(R)$
(for an application, see section~\ref{s_galaxy}).

Let $\sigma_R(R)$ be the given radial velocity dispersion profile which we 
want to fix, where $R$ is the cylindrical radius. After each
evolutionary step (see Sect.\ref{generalidea}) we need to change
slightly the radial velocities of particles in order to fix this profile.
The model is divided into  $n_{\rm div}$ concentric cylindrical
annuli, each containing the same 
number of particles. For each annulus $j$, we calculate the target value of
the radial velocity dispersion 
\begin{equation}
\sigma_R^j = \sigma_R(R_j) 
\, ,
\end{equation}
where $R_j$ is the mean value of the $R$ coordinate of all particles
in part $j$. 
The new radial velocity of the $i$-th particles in the $j$ region is
then prescribed as follows.
\begin{equation}
v_{R i} = v_{R i}^\prime \, \sigma_R^j / \sigma_R^{j \prime}
\, ,
\end{equation}
where $v_{R i}^\prime$ is the current value of
the $i$-th particle radial velocity,
$v_{R i}$ is the corrected $i$-th particle radial velocity and 
$\sigma_R^{j \prime}$ is the current value of radial velocity dispersion in
part $j$. We note that in this scheme we have assumed that the mean
radial velocity is equal to zero.

\subsubsection{Fixing the radial anisotropy profile}
\label{s_fixrasov}

This algorithm is very similar to previous one and is very useful for
building spherical models with a given profile of velocity
anisotropy. We 
will use it for building the halo model in section~\ref{s_galaxy} and
it has also been used in \citet{SR08}.

Let $\sigma_{\theta}$, $\sigma_{\varphi}$ and
$\sigma_r$ be the velocity dispersions in the $\theta$, $\varphi$ and $r$
directions of spherical coordinate system and let us aim e.g. for a
model with a given profile, $\beta(r)$, of the
$\sigma_{\theta}/\sigma_{r}$ ratio. The model is divided in concentric
spherical shells, each containing the same number of particles. 
For each shell $j$, we calculate the target value of
$\sigma_{\theta}/\sigma_{r}$ 
\begin{equation}
\beta_j = \beta(r_j) \, ,
\end{equation}
where $r_j$ is the mean value of the $r$ coordinate of all particles
in shell $j$. We will attempt to obtain this ratio by changing
appropriately the $\theta$ component of particle velocities
(alternatively, we could have changed the $r$ component).
The new $\theta$ velocity component of the $i$th
particle in the $j$th region will then be prescribed by
\begin{equation}
v_{\theta i} = v^{\prime}_{\theta i} \, \beta_j \,
\displaystyle\frac{\sigma^{j \prime}_{r}}{\sigma^{j \prime}_{\theta}}
\, ,
\end{equation}
where $v_{\theta i}^\prime$ and $v_{\theta i}$ are, respectively, the
current and the corrected values of
the $i$-th particle $\theta$ velocity component and 
$\sigma^{\prime}_{r}$ and $\sigma^{\prime}_{\theta}$ are 
the current values of the radial and $\theta$ velocity dispersion,
respectively, in part $j$. 

\subsubsection{Fixing the line-of-sight mean velocity or the 
line-of-sight velocity dispersion in the case of
an edge-on disk}
\label{s_fixvlos}
 
In this section, we describe two algorithms, one for fixing the
line-of-sight 
mean velocity and the other for fixing the line-of-sight dispersion of
an axisymmetric disk.
To set the notation, let us assume that the stellar disk
rotates about the $z$-axis, the disk plane lies in the $(x,y)$ plane and
the line of sight is along the $y$ axis (edge-on disk). 
We invert the sign of $v_y$
for each particle with 
$x<0$ in order to make the half disk with $x<0$ kinematically
identical to the half with $x>0$ and flip the $x < 0$ particles on the
$x > 0$ part. We then 
divide the disk in slits parallel to the $(y,z)$ plane and at different
distances from the centre, i.e. at different values of $x$, in such a
way that all slits have the same number of particles. 

Let us denote by $\bar v_{\rm los}(x)$ the desired profile of the
line-of-sight mean velocity, i.e. the mean value of $v_y$ after
integration along the line of sight. For each slit $j$ we calculate the
target value of the line-of-sight mean velocity 
$\bar v_{\rm los}^j = \bar v_{\rm los}(x_j)$  
(where $x_j$ is the mean value of $|x|$ for particles in part $j$) and 
the current value of the line-of-sight mean velocity 
$\bar v_{\rm los}^{j\prime}$ (as the mean value of $v_y$ for all
particles in slice $j$). The new $y$ velocity 
component of particle $i$ in region $j$ should then be
\begin{equation}
v_{y i} = v_{y i}^{\prime} + (\bar v_{\rm los}^j - \bar v_{\rm los}^{j\prime}), 
\end{equation}
where $v_{y i}^\prime$ is the current value of
the $i$-th particle $y$ velocity and
$v_{y i}$ is the corrected $i$-th particle $y$ velocity. 
Particles which were flipped to $x>0$ part have to be flipped back and
the sign of their $y$ velocity component reversed.
The particles are then azimuthally mixed to make the velocity
distribution axisymmetric and the step is concluded.
Of course, in this way we have tampered with $v_{los}$, but this
unavoidable. Nevertheless, after a number of iterations, both the
axisymmetry and the desired $v_{los}$ will be achieved. 

The algorithm for fixing $\sigma_{\rm los}(x)$ is very similar, except
that we have to calculate in each slit the current value of the 
line-of-sight velocity dispersion $\sigma_{\rm los}^{j\prime}$ as the
dispersion of $v_y$ for all particles in slit $j$. Let
 $\sigma_{\rm los}(x)$ be the desired profile of the
line-of-sight velocity dispersion. In order to achieve this, the new
$y$ velocities should be modified as follows
\begin{equation}
v_{y i} = (v_{y i}^{\prime} - \bar v_{\rm los}^{j\prime})  
\frac{\sigma_{\rm los}^{j}}{\sigma_{\rm los}^{j\prime}}
+ \bar v_{\rm los}^{j\prime},
\end{equation}
where $\bar \sigma_{\rm los}^j = \sigma_{\rm los}(x_j)$ is the target
value of the line-of-sight velocity dispersion.
We note that in this algorithm we have to take into account that 
the value of $\bar v_{\rm los}^{j\prime}$ need not necessarily be
equal to zero. As previously, we still have to 
flip back particles which were flipped to the $x>0$ part,
invert the sign of their
$y$ velocity component with and mix the particles azimuthally. 
 
\subsubsection{Fixing velocity isotropy conditions}
\label{s_isotropy}

An isotropic velocity distribution depends only on the velocity module
and not on the direction of the velocity. Our algorithm for
 fixing it is very
simple. For each particle, we keep the velocity module unchanged and
randomise its direction, thus ensuring that the velocity distribution
is isotropic. For spherical isotropic models, the distribution
function (DF) is known, at least formally, or numerically. Thus 
the construction of such models can be considered as a test of
the iterative method, and we have verified in a number of cases that
the models constructed by the iterative
method are identical to the models constructed by using known equilibrium
DF. The construction of spherical
isotropic models with the iterative method was first described in
\citet{RS06}. That work, however, used an old and rather complicated
algorithm for transferring the velocity distribution. Here we use a
different, superior algorithm, based on the description in
Sect.~\ref{s_transvel}. We again 
made sure that the models thus constructed are identical to the models
obtained by using a known DF. Furthermore, we also used this 
algorithm for constructing models which are not fully
isotropic models, but rather not-very-far from isotropic (see
section~\ref{s_3axial} and \ref{s_galaxy}). 

%\subsubsection{Fixation of total angular momentum about z-axis}
%\label{s_fixLz}

%Let $L_z$ be given value of angular momentum,
%$L_z^\prime$ be the current value of angular momentum.
%We need slightly change the azimuthal velocities of particles to make
%angular momentum equal $L_z$.
%New azimuthal velocities of particles were prescribed
%as follows
%\begin{equation}
%v_{\varphi i} = v_{\varphi i}^\prime +
%\frac{(L_z - L_z^\prime)}{{\sum_{j=1}^{N}R_j m_j}}
%\, ,
%\end{equation}
%where $v_{\varphi i}^\prime$ is the current value of
%the $i$-th particle azimuthal velocity,
%$v_{\varphi i}$ is the corrected $i$-th particle azimuthal velocity.

%This algorithm was used in \citet{RS06} and \citet{RO08}.

\subsection{How many parameters should we fix?}
\label{s_howmany}

  The goal of the iterative method is to construct equilibrium $N$-body models
with given parameters (i.e. with a given mass distribution and with
given kinematic constraints). There are in general three possible cases
with respect to the number of constraints.

In the first case the number of given parameters, or constraints are
such that only one equilibrium model can exist. In this case, we can
expect that the iterative method will converge to 
this unique equilibrium model, independent of the initial state from
which the iteration is started. For example, it is
known that for a spherical model with a given mass distribution 
only one isotropic equilibrium DF exists. If we construct a spherical model
by using the iterative method and we fix velocity isotropy as a
kinematical constraint
(section~\ref{s_isotropy}), then the iteration always converges to the
same model, independent of the initial state, as expected.

In the second case, the number of give parameters is such that many
equilibrium models can exist, i.e. this number is insufficient for
determining uniquely the equilibrium model. In this case we can expect
that the result of the iterative method will depend on the choice of the
initial model. The iteration will converge to the equilibrium model
which is ``nearest'' in some sense to the initial model. Alternatively, the 
iteration will converge to some specific, in some sense, model.
For example, when constructing a triaxial model in section~\ref{s_3axial} 
we fix only the mass distribution and do not set any 
kinematic constraints. Of course in this case the result of the
iterations will depend on the initial state (see
section~\ref{s_3axial} for details). Another, more involved, example
is the construction of a disk model with given total angular
momentum. In principle, many such equilibrium models are possible, yet
the iterations of \citet{RO08} always converged to the same
model. It is unclear why this is the case, but it could be due to a
specificity of the model (see \citealt{RO08} for details). 
  
The last possibility is that for the adopted parameters, no
equilibrium model exists, i.e. we have fixed too many parameters. 
In this case the iteration will either not converge at all, or it will
converge to a system with the parameters we have fixed, which is in
non-equilibrium, but close in some sense to equilibrium.

\subsection{Technical comments}
\label{s_tech}

In this section we elaborate a few important technical points, useful
for anybody wishing to apply the iteration method.

One of the free parameters of the iterative method is the duration
$t_i$ of each iteration, i.e. 
the time interval over which the system is evolved during each 
iteration. How should the numerical value of $t_i$ be chosen? It is
clear that this time should not be too short, so as to allow
the system to evolve sufficiently during one iteration step.
On the other hand, it should not be too long either, so as not to
permit the evolution of various instabilities;
otherwise, these instabilities may change the system substantially.
For example, when constructing a disk system it is necessary to use
iteration steps considerably shorter that the growth time of the bar
instability, which of course varies strongly from one model to
another. For this reason, there is no strict criterion and $t_i$ should be
chosen empirically. Our experiments have shown that it is usually
better to try relatively big $t_i$ values, thus ensuring a much faster
convergence. Moreover, in some situations the iterations for 
relatively small
$t_i$ don't converge at all, while iterations for relatively big $t_i$
do. This was, for example, the case when we constructed a model with 
relatively cold stellar disk. So if 
iterations don't converge or they converge too slowly, it is often
useful to consider bigger $t_i$ (within of course reasonable
limits). Examples of appropriate $t_i$ values are given in all
examples in Sect.~\ref{s_models}. Moreover,
our simulations have shown that, if we take $t_i$
within reasonable limits, the result of the iteration is 
the same (within the noise limits) and independent of $t_i$, provided
of course the chosen 
number of parameters and constraints allow a single solution. If the
latter is not the case, and the result of the iteration depends on its
starting model, then of course the result can depend also on 
$t_i$. 
 
Another parameter of the iterative method is the parameter $n_{nb}$ in the
algorithm of velocity transfer (see section~\ref{s_transvel}). Its
value has been chosen in a more or less ``ad hoc'' manner and, by
trial and error, we have
found that a value of $n_{nb}=10$ is often satisfactory.
Our test simulations have shown that the results of the
iterative method for $n_{nb}=10$ and $n_{nb}=100$ are practically the
same, at least for a total number of particles as those used in our
trials, i.e. of the order of a few hundred thousands to a couple of
millions.

The most computer costly part of our method is the computing of the 
evolution of the system in each iteration, since
the computing cost of all other parts of the method is very small. For
this reason it is recommended to use a fast $N$-body code and we have
adopted gyrfalcON \citep{D00, D02}. Furthermore, our test simulations  
have shown that computation of the evolution can be carried out with 
relatively low accuracy. This is mainly due to the fact that we need
to calculate the evolution only during short time intervals, so that
errors do not have sufficient time to accumulate. Therefore, in the iterative
method we use GyrfalcON with relatively big values of the tolerance
parameter $\theta_t$ and of the time-step (see section~\ref{s_models}). 
Usually the total computing cost is considerably smaller, but still of the
same order as that necessary to run a simulation with the constructed
model. 
This of course will depend on whether we can start the
iterations form a model reasonable close to the final one, or whether
lack of any a priori knowledge leads us to start, e.g. from zero
initial velocities. 
A `trick' which helps reducing the computing
cost is to make a few iterations initially with a small number $N$ of
particles and then gradually increase $N$ to the required number.
In the procedure of velocity transfer described in section~\ref{s_transvel}
the number of particles in the old and the new system can be
different. So in the next
iteration step we can get a system with a larger number of particles.

\section{Examples of models}
\label{s_models}
 In this section we consider three examples of models constructed by
our method, namely a triaxial model of a spheroid, a multi-component
model of a disk galaxy and a model with given line-of-sight kinematics.  

\subsection{Triaxial model}
\label{s_3axial}

Our first example is that of a triaxial model. As mass distribution we use a
Plummer sphere flattened in two dimensions. 

\begin{equation}
\label{eq_plum_ro}
\rho(x,y,z) = \frac{3 M_{\rm pl}}{4 \, \pi \, a \, b \, c} \, {a_{\rm pl}^2}
\left( \displaystyle\frac{x^2}{a^2} + \frac{y^2}{b^2} + \frac{z^2}{c^2} 
+ a_{\rm pl}^2\right)^{-5/2} \, ,
\end{equation}

\noindent
where $M_{\rm pl}$ and $a_{\rm pl}$ are the total mass and the scale-length of
the model and $a$, $b$, $c$ are rescaling parameters. In the present
specific example we
discuss a model with the following parameters: $M_{\rm pl} = 1$, 
$a_{\rm pl} = 3 \pi / 16$, $a = 1$, $b = 0.8$ and $c = 0.7$. 
Scaling our
model to an elliptical galaxy with $a=3$~kpc and 
$M_{\rm pl} = 10^{11} \; {\rm M}_{\odot}$, gives a time unit
$t_u \approx 17$~Myr.

Our target is to construct an equilibrium $N$-body system with this
given mass distribution. We didn't impose  any well-defined  
restriction on the kinematics of the system, but aimed instead for a
velocity distribution not very far from isotropic. Our initial model was
cold with velocities equal to zero. Our target model was reached in
$50$ iterations. 
For the first $10$ we fixed a condition of velocity isotropy 
(section~\ref{s_isotropy}), while for the last $40$ we didn't fix any kinematic
parameters. If we performed all $50$ iterations without fixing any kinematic
parameters, we would also obtain an equilibrium model but with a higher
degree of anisotropy. We chose $N=500\,000$ and an iteration time
$t_i=10$. The integration step and the softening length were taken
$dt=1 / 2^7$ and $\epsilon=0.01$, respectively. The tolerance 
parameter for gyrfalcON was set to $\theta_t=0.9$ (see
section~\ref{s_tech}) and we used the
``transvel\_3d'' modification of the algorithm of velocity transfer (see
section~\ref{s_transvel}).

\begin{figure}
\begin{center}
\includegraphics[width=5.9cm,angle=-90]{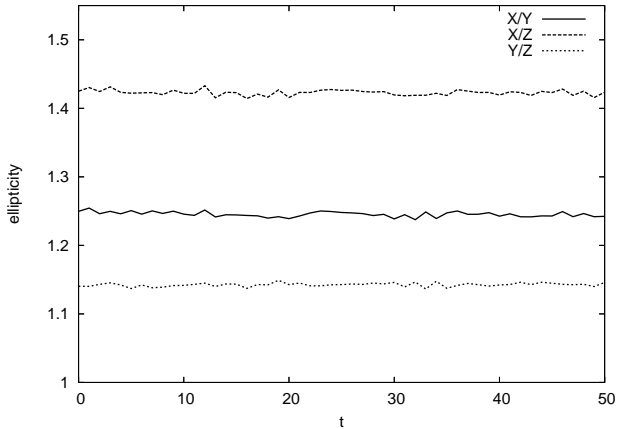}
\end{center}
\caption{ 
%Example of a triaxial model constructed using the iterative
%method. The upper snapshots show three different projection of the model. 
%The bottom panel shows the evolution of axial ratios of the models for
%the three projections this demonstrating that they do not evolve,
%i.e. that the model is very equilibrium. 
 Evolution of the ellipticity for the three projections of the
   triaxial model constructed with our iterative method.
   Note that they do not evolve, i.e. that the model is in equilibrium. 
}
\label{fig_3d}
\end{figure}

In such a case it is crucial that the final model be sufficiently
close to equilibrium so that the axial ratios do not tend to unity
after some time evolution, as it happens for many other techniques
used in calculating triaxial equilibria. To test this we evolved our
model for 50 time units, i.e. roughly 50 crossing times for the
scale-length of the system $a_{\rm pl}$. The integration step and
softening length were taken $dt=1 / 2^8$ and $\epsilon=0.005$,
respectively -- in agreement with the recommendations of \citet{RS05} (see
also \citealt{AFLB00}) -- and the tolerance
parameter for gyrfalcON was set to $\theta_t=0.6$.
Fig.~\ref{fig_3d} shows the evolution of the ellipticity of
the model for three different projections. This was  
calculated as the ratio of the medians of the absolute values of 
corresponding particle coordinates. As can be seen, the shape of the
model is practically unchanged during the evolution. We also made sure
that the model also
conserved all its other properties, thus demonstrating that 
it is indeed very close to equilibrium.

\subsection{Multi-Component model of a disk galaxy}
\label{s_galaxy}

As a second example, we constructed a model of a disk galaxy
consisting of three components: a stellar disk with a given
profile of radial velocity dispersion, a non-spherical bulge and a halo with
a given anisotropy profile. 

To start, we need to define the mass distribution in each of the components.
The disk model is an exponential disk with density:

\begin{equation}
\label{eq_disk}
\rho_{\rm d}(R,z)=
\displaystyle\frac{M_d}{4 \pi R_d^2 z_0}
\exp\left(-\frac{R}{R_d}\right)
{\rm sech}^2\left(
\frac{z}{z_0} \right) \, ,
\end{equation}

\noindent
where $M_d$ is the total disk mass, $R_d$ is the disk scale length, $z_d$ is
its scale height and $R$ is the cylindrical radius.
The halo model is a truncated NFW halo (\citealt{N96}) 

\begin{equation}
\rho_{\rm h}(r) = C_{\rm h} \frac{exp(-r^2/r_{\rm th}^2)}
{(r/r_h)(1+r/r_h)^2} \, ,
\label{eq_NFW}
\end{equation}

\noindent
where $r_h$ is the halo scale length, $C_h$ is a parameter defining
the mass of the halo and $r_{\rm th}$ is the truncation radius of the halo. For
the bulge we used a truncated and flattened Hernquist sphere (\citealt{H90}) 
with density

\begin{equation}
\rho_{\rm b}(R, z) = \frac{M_b^{\prime} r_b}{2 \pi q} 
\frac{exp(-d^2/r_{\rm tb}^2)}{d (d + r_b)^3} \, ,
\end{equation}

\noindent
where
\begin{equation}
d = \sqrt{R^2 + \frac{z^2}{q^2}} \, ,
\end{equation}
$r_b$ is the bulge scale length, $M_b^{\prime}$ is the total bulge mass before
truncation, $r_{\rm tb}$ is the truncation radius of the bulge and 
$q$ is the flattening parameter. 

\begin{figure}
\begin{center}
\includegraphics[width=5.9cm,angle=-90]{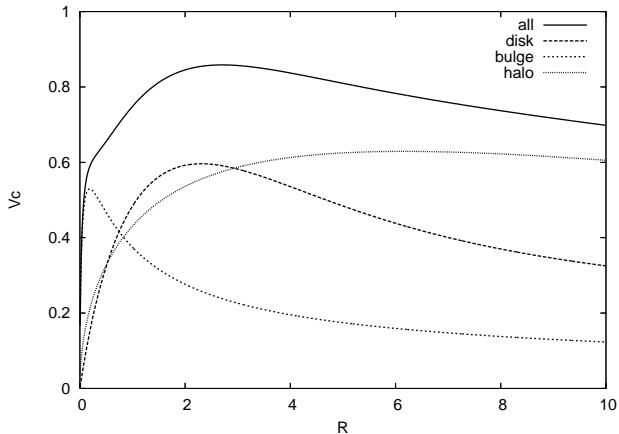}
\end{center}
\caption{Rotation curve for the disk galaxy model. The solid line is the
total rotation curve. We also show the contributions from the disk, halo and
bulge components.}
\label{fig_galaxy_vc}
\end{figure}

\noindent
For the parameter values we chose for the 
disk: $M_d = 1$, $R_d = 1$, $z_0=0.2$,
for the halo: $r_h=4$, $C_h=0.01$, $r_{\rm th}=14$
and for the bulge: $r_b=0.2$, $M_b^{\prime}=0.2$, $r_{\rm tb}=2$, $q=0.7$.
For these parameters the total mass of the bulge and of the halo are
$M_b \approx 0.15$ and $M_h \approx 4.98$, respectively. 
We use units such that the constant of gravity is $G=1$. Scaling our
model to a disk galaxy with $R_d=3.5$~kpc and $M_d= 5 \cdot 10^{10} \;
{\rm M}_{\odot}$, gives a time unit
$t_u \approx 13.8$~Myr and a velocity unit $v_u \approx 247.9 \; {\rm km/s}$.
The rotation curve for our model is shown in
Figure~\ref{fig_galaxy_vc}, which also displays the contribution of the 
disk, halo and bulge components separately.

\begin{figure}
\begin{center}
\includegraphics[width=5.9cm, angle=-90]{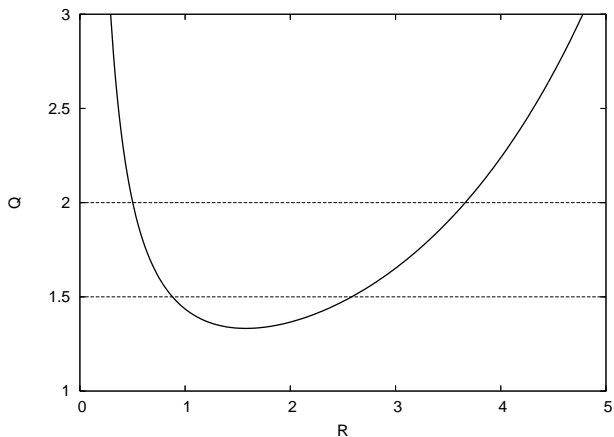}
\end{center}
\caption{Radial profile of the Toomre parameter Q for the disk in the
  model described in Sect.~\ref{s_galaxy}.}
\label{fig_galaxy_Q}
\end{figure}

To make the exercise more realistic, we still need to choose
kinematical constraints for each of the components, although, as we
have already mentioned, these are not obligatory for our method.
We created the disk with the following profile

\begin{equation}
\label{eq_svR}
\sigma_R(R) = 0.3 \cdot \exp\left(-R/3\right) + 0.2 \cdot
\exp\left(-R/0.5\right) \, ,
\end{equation}
where $\sigma_R$ is the radial velocity dispersion. From the mass
model of the galaxy and from profile~(\ref{eq_svR}) we can calculate
the radial profile of the Toomre parameter Q (\citealt{T64}), shown 
in figure~\ref{fig_galaxy_Q}.
Our main target here is to demonstrate that our method can construct an
equilibrium model of the disk with any realistic profile of $\sigma_R(R)$.
The choice of profile in eq.~(\ref{eq_svR}) is more or less arbitrary,
but demonstrates that
our method can work with more elaborate profiles than a
single exponential function.

 When constructing the bulge, we did not impose any specific kinematic
constraints. Instead, we aimed for a model not-very-far from
isotropic, as in the case of the triaxial model of Sect.~\ref{s_3axial}.

For the halo we adopted a velocity anisotropy profile, so as to test a
different kind of constraint. More specifically we chose

\begin{equation}
\label{eq_rasov_halo}
\frac{\sigma_{\theta}(r)}{\sigma_{r}(r)} =
\frac{0.2}{\displaystyle\sqrt{\left(\frac{r}{0.9}\right)^2+1}} + 0.8 \, ,
\end{equation}
where $\sigma_{\theta}$ and $\sigma_{r}$ are the velocity dispersion
in the $\theta$
and the $r$ direction in a spherical coordinate system. Note that we
constructed the phase model of the halo in the presence of the 
non-spherical potential of the disk
and bulge, i.e. we don't have spherical symmetry and in the halo
equilibrium model 
$\sigma_{\theta}$ should not be equal to $\sigma_{\varphi}$ (the velocity
dispersion in $\varphi$ direction). So when we constructed the halo model we
fixed only the fraction $\sigma_{\theta}/\sigma_{r}$, but did not fix
the fraction
$\sigma_{\varphi}/\sigma_{r}$. This is different from the case
of the isolated spherical NFW halo, constructed in 
\citet{SR08}. And again we want to underline that kinematical constraints
are not obligatory. We could also construct the halo, or the bulge,
without any kinematical constraints, or with another type of
kinematical constraints. For example 
we can construct a rotating halo with given total angular momentum or
a rotating halo with a given profile of the mean azimuthal velocity. 

Once the mass models and the required kinematical parameters for each of
three components are defined, we can apply the iterative method for
constructing an equilibrium $N$-body model for the whole system. In this
specific example we chose $N_d=200000$, $N_b=30324$, $N_h=995978$ for
the number of particles in disk, bulge and halo,
respectively. With these numbers, the mass of particles in all components is
the same. We constructed each of these components separately in the
rigid potential of all other components. In order to take into
account the external potential, we need to make only one small evident 
modification of the iterative method. Namely, when we need calculate the
evolution of the system during the iteration time (see
fig.~\ref{fig_scheme}), we simply need to do it in the presence of the
external potential. This can be done either by introducing an
analytical external potential to the gyrfalcON program, or we can add
it as a rigid $N$-body system. In current
work we use the latter. For example, in order to add a rigid halo we simply
add in the system 
rigid particles according to the mass distribution of the halo.

Let us first describe the disk construction.
Our initial model was a cold disk where all particles move on
circular orbits. Indeed, the circular velocity can be easily
calculated, since the mass distribution in the model is known. Had we,
instead, started off with zero disk velocities, we would have again obtained an
equilibrium model, but with counter rotating subsystems. This
will happen because we fix only the profile of $\sigma_R(R)$ and do not fix any
parameters defining the direction of rotation. It is therefore better
to start off the iterations with a rotating disk, unless of course a
disk with counter-rotating components is specifically
sought. Note that the result of the iteration is
independent of
the initial iterative guess for the disk rotation. For example it will be
the same if initially all the disk particles have tangential velocities equal
to half of the circular velocity, or twice that. 

We made $50$ iterations, each with $t_i=20$. The integration step and
softening length were taken $dt=1 / 2^4$ and $\epsilon=0.04$,
respectively and the tolerance
parameter for gyrfalcON was set $\theta_t=0.9$. In order to fix the
$\sigma_R(R)$ profile we used the
algorithm described in section~\ref{s_fixsvR_R}. The number of layers
in this algorithm was $n_{\rm div}=200$ (see section~\ref{s_fixsvR_R}).  
We used the
``transvel\_cyl'' modification of the algorithm of velocity transfer (see
section~\ref{s_transvel}). This algorithm also was used for constructing
the bulge and halo components. We call this disk model 
DISK.SVR.

 For constructing the bulge we also made $50$ iterations. The other
 parameters 
for this construction were  $t_i=10$, $dt=1 / 2^6$, $\epsilon=0.02$ and
$\theta_t=0.9$.
Our initial model was a cold model with velocities equal to zero.
During the first $10$ iterations we fixed a condition of velocity isotropy
(section~\ref{s_isotropy}), while we did not set any kinematical
constraints during the last $40$ iterations.

To construct the halo we used again 50 iterations and the remaining
parameters were taken as follows :
$t_i=50$, $dt = 1 / 2^4$, $\epsilon=0.04$ and
$\theta_t=0.9$. For fixing the velocity anisotropy radial profile
(\ref{eq_rasov_halo}) we used the algorithm described in
section~\ref{s_fixrasov}. The number of layers in this algorithm was 
$n_{\rm div}=500$.

\begin{figure*}
\begin{center}
\includegraphics[width=14cm]{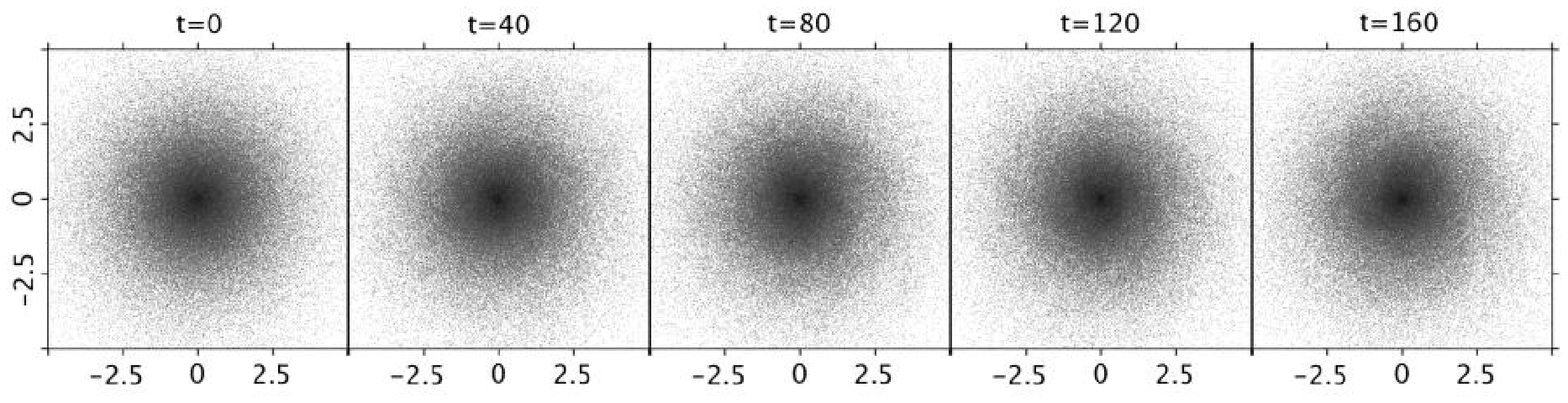}
\includegraphics[width=14cm]{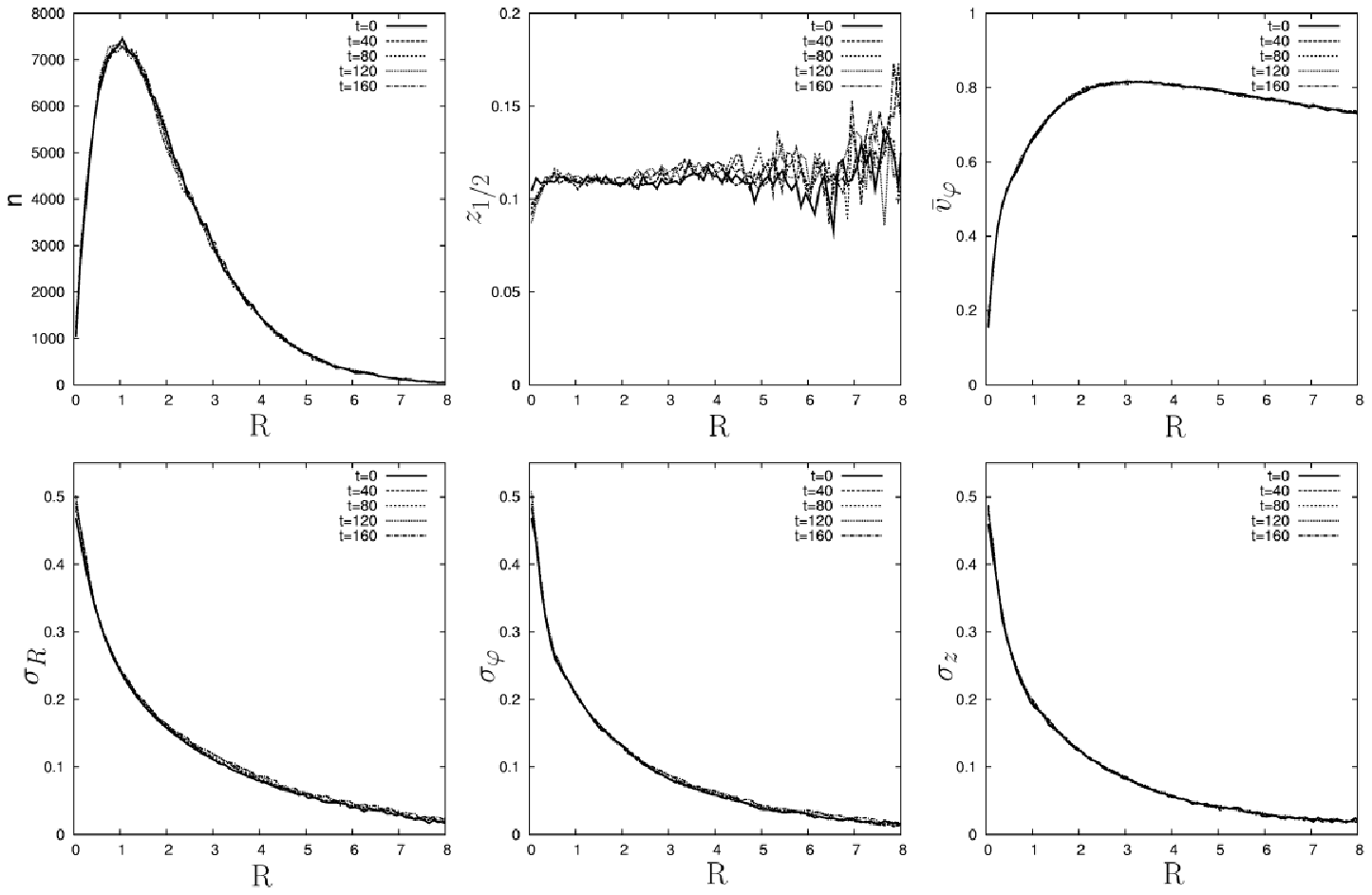}
\end{center}
\caption{Initial evolutionary stages for the disk of the constructed 
disk galaxy model. The evolution of the model was calculated with live
disk, halo, and bulge components (see also fig.~\ref{fig_bulge} and
~\ref{fig_halo}). From left to right, the upper 
snapshots show the disc views face-on for times 0, 40,
80, 120 and 160 and the grey scales are logarithmically spaced. 
The middle and bottom panels show the
dependence of various disc quantities on the cylindrical radius $R$ at
the same times. Here $n$ is the number of particles in concentric
cylindrical layers; $z_{1/2}$ is the median of the value $|z|$, i.e a
measure of the disc thickness (see \citealt{SR06}) and
$\bar v_{\varphi}$, $\sigma_R$, $\sigma_{\varphi}$ and $\sigma_z$ 
are four moments of the velocity distribution. At the beginning of the
evolution ($t$ = 0) the disk  has, by construction, the radial
dispersion profile given by eq. (\ref{eq_svR}). 
}
\label{fig_disk}
\end{figure*}

\begin{figure*}
\begin{center}
\includegraphics[width=14cm]{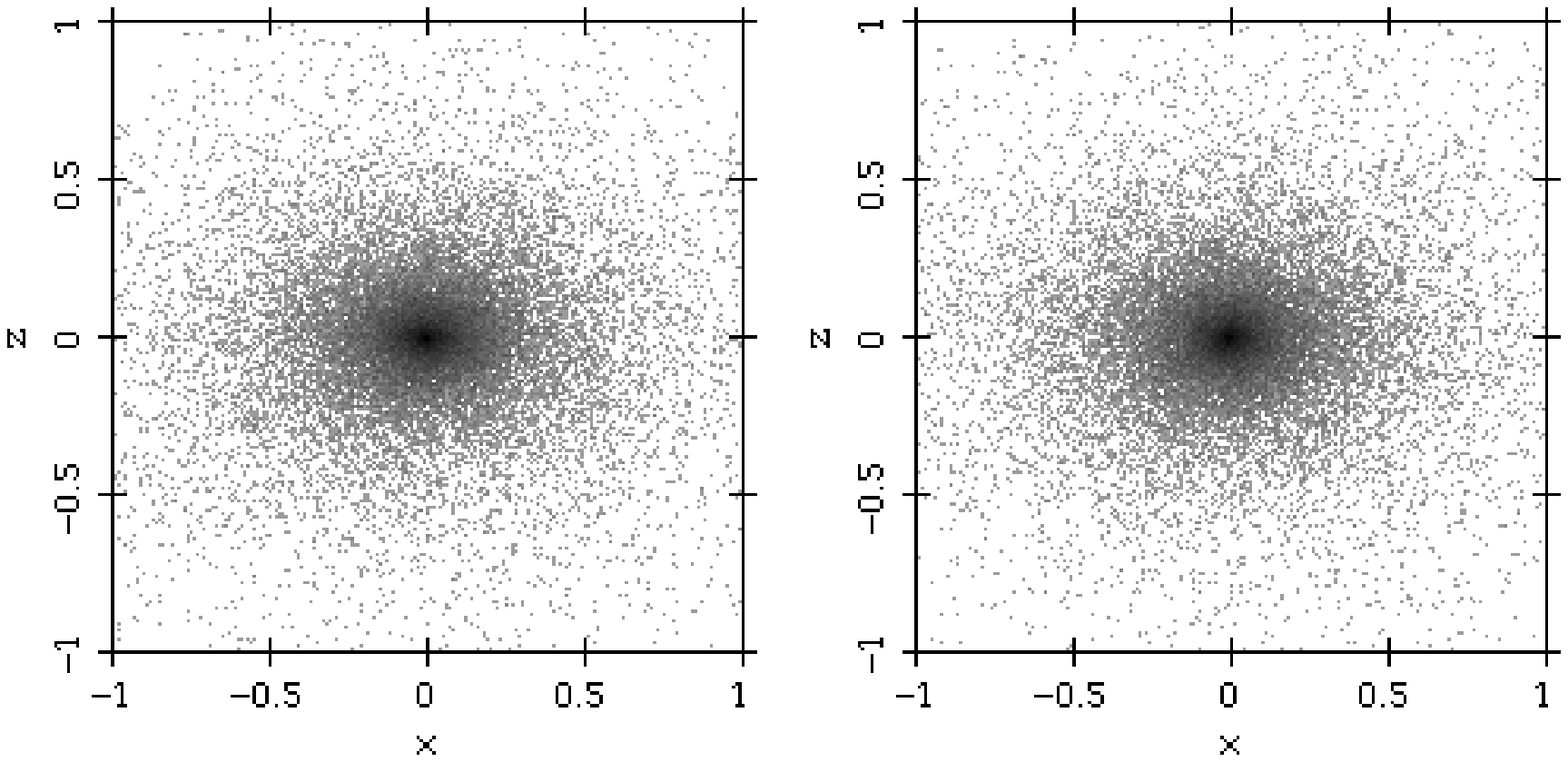}
\includegraphics[width=14cm]{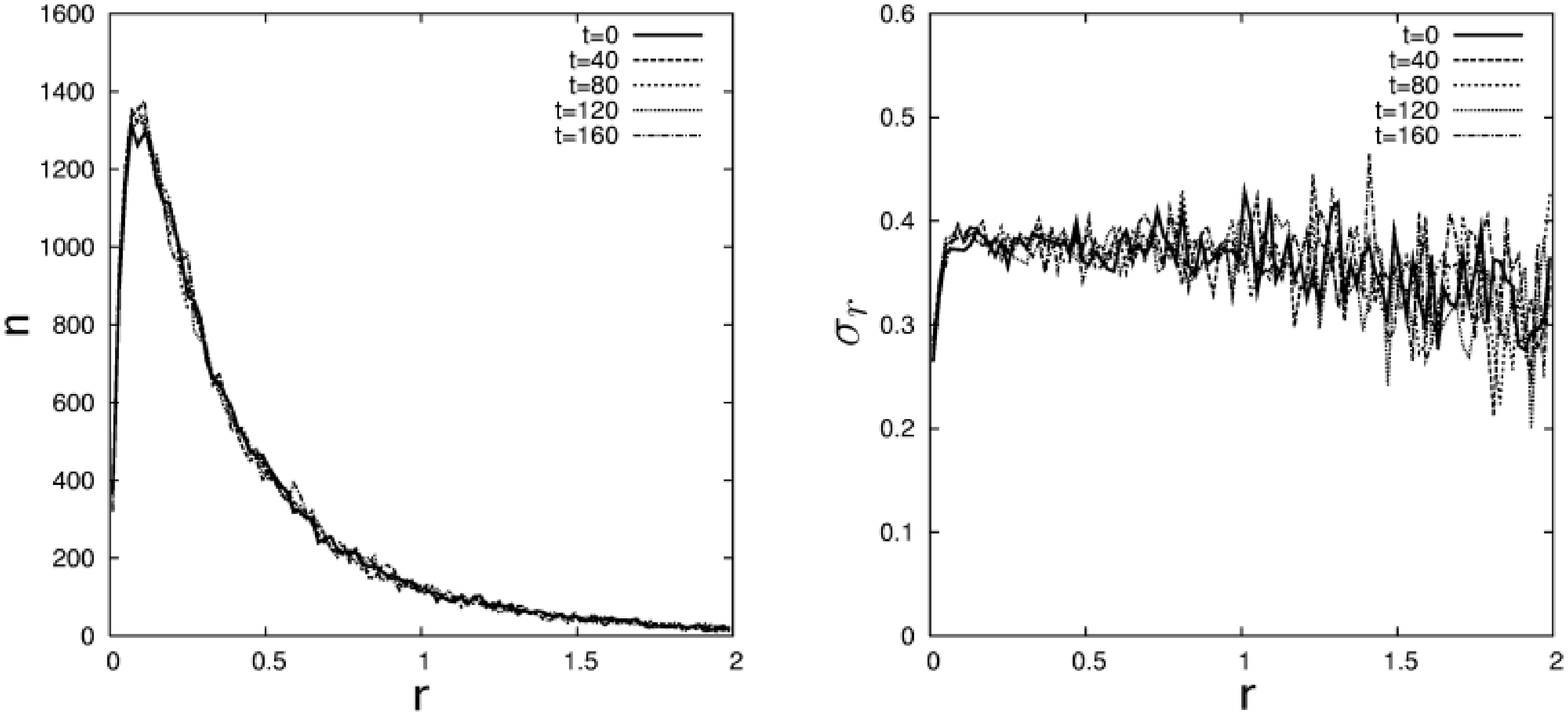}
\end{center}
\caption{Initial evolutionary stages for the bulge of the constructed
  disk galaxy model. The evolution of the model was calculated with live disk, halo,
and bulge (see also fig.~\ref{fig_disk} and ~\ref{fig_halo}). The upper
snapshots show the bulge viewed edge-on for two moments of time (0, 160); 
the grey intensities correspond to the logarithms of
particle numbers in the pixels. The bottom panels show the
dependence of two parameters of the bulge on the spherical radius $r$ 
for various moments of time. Here $n$ is the number of particles in concentric
spherical layers; $\sigma_r$ is the velocity dispersion in the $r$
  direction (in spherical coordinate system). We demonstrate these
  parameters in order to show
that the model is close to the equilibrium. For astrophysical applications it
should be taken into account that the bulge in our model is not spherical.}
\label{fig_bulge}
\end{figure*}

\begin{figure*}
\begin{center}
\includegraphics[width=14cm]{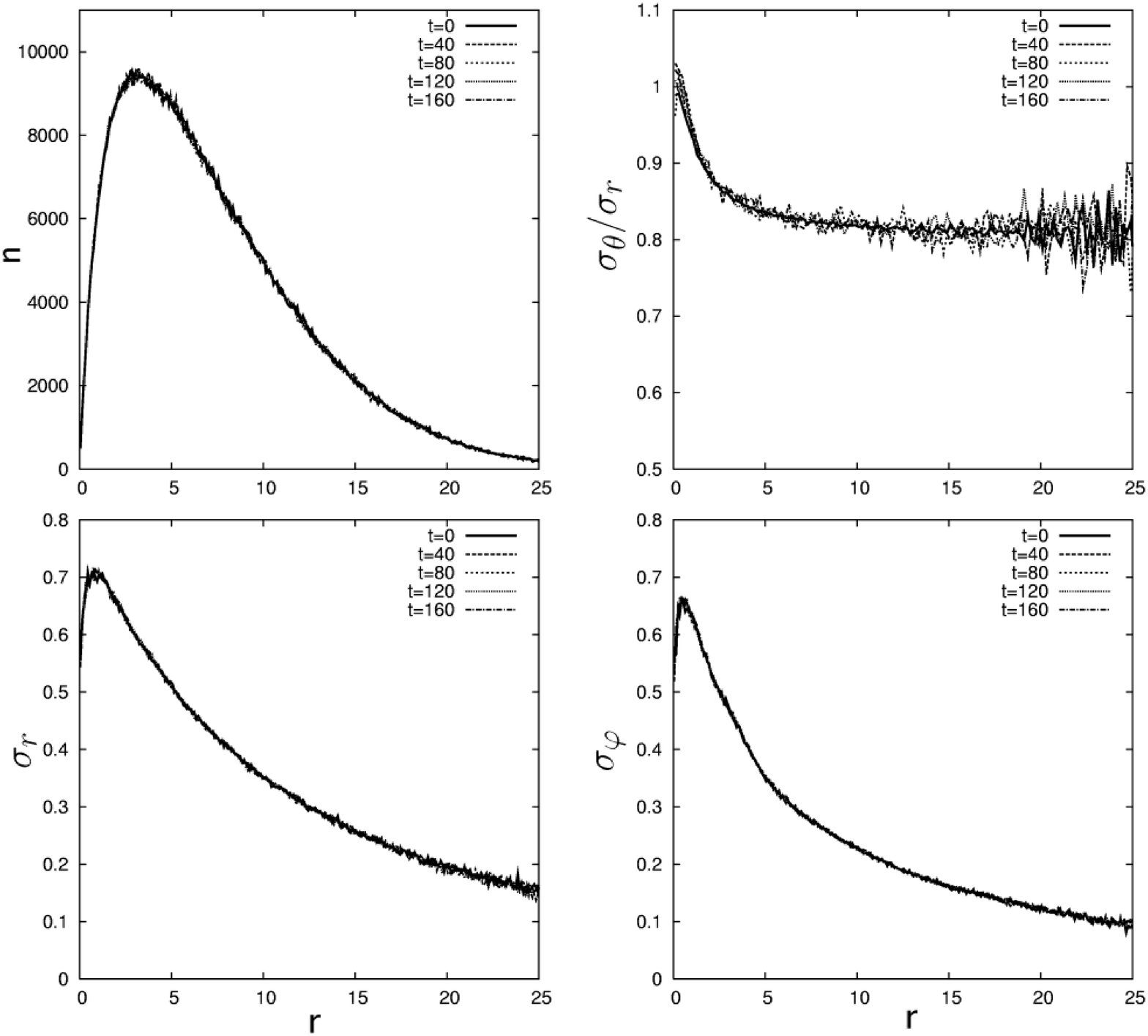}
\end{center}
\caption{Initial evolutionary stages for the halo of the constructed
disk galaxy model. The evolution of the model was calculated with
live disk, halo, and bulge components (see also fig.~\ref{fig_disk}
and ~\ref{fig_bulge}). We show the 
dependence of various halo parameters on the spherical radius $r$ for
various moments of time. 
Here $n$ is the number of particles in concentric
spherical layers (upper left panel); $\sigma_{\theta}/\sigma_{r}$ is the
ratio of velocity dispersion in the $\theta$ and $r$ directions (upper right panel);
$\sigma_r$ is velocity dispersion in the radial direction (bottom left panel);
$\sigma_{\varphi}$ is the velocity dispersion in the $\varphi$
  direction (bottom right panel). At the initial moment of time
the halo has the profile of $\sigma_{\theta}/\sigma_{r}$ given by (\ref{eq_rasov_halo}).}
\label{fig_halo}
\end{figure*}

Once all three components of our model were constructed, we simply
stacked them in order to obtain the complete system. In order to check
whether this was indeed near equilibrium, as it should, we simply
evolved with a full $N$-body simulation, using again gyrfalcON, now
with an integration step and
softening length of $dt=1 / 2^7$ and $\epsilon=0.02$ (parameters were
chosen according to recommendations of \citet{RS05}). The tolerance
parameter for gyrfalcON was set $\theta_t=0.6$. The evolution of the
total system over 160 time units is illustrated separately for
the disk, bulge and halo components in figures~\ref{fig_disk},
\ref{fig_bulge} and \ref{fig_halo}, respectively. 
These show that all three components of the constructed model
conserve their structural and dynamical properties very well,
demonstrating that the constructed model is indeed close to
equilibrium as it should.

An interesting question arises
in connection with the equilibrium of our model: 
how well do the moments of the
velocity distribution in the constructed disk satisfy the 
equilibrium Jeans equations (see \citealt{BT87})?

\begin{equation}
\label{eq_jeans}
\left\{
\begin{array}{rcl}
{\overline v}_{\varphi}^2 &=& v_{\rm c}^2 + \sigma_R^2  -
\sigma_{\varphi}^2 + \displaystyle \frac{R}
{\rho_{\rm d}}
\frac{\partial (\rho_{\rm d} \sigma_R^2)}{\partial R} \, , \\
\sigma_{\varphi}^2 &=& \displaystyle\frac{\sigma_R^2 R}
{2 \overline v_{\varphi}}
\left( \frac{\partial \overline v_{\varphi}}{\partial R} +
\frac{\overline v_{\varphi}}{R} \right) \, , \\
\displaystyle\frac{\partial (\rho_{\rm d} \sigma_z^2)}{\partial z} &=&
-\rho_{\rm d}
\displaystyle\frac{\partial \Phi_{\rm tot}}{\partial z} \, .\\
\end{array}
\right.
\end{equation}
Here $\Phi_{\rm tot}$ is the potential generated by all the 
components of our model
(disk, halo and bulge), $\bar v_{\varphi}$, $\sigma_R$, 
$\sigma_{\varphi}$ and $\sigma_z$ 
are four moments of the velocity distribution in the disk (mean azimuthal
velocity and velocity dispersions in the $R$, $\varphi$ and $z$ directions,
respectively).

\begin{figure*}
\begin{center}
\includegraphics[width=17cm]{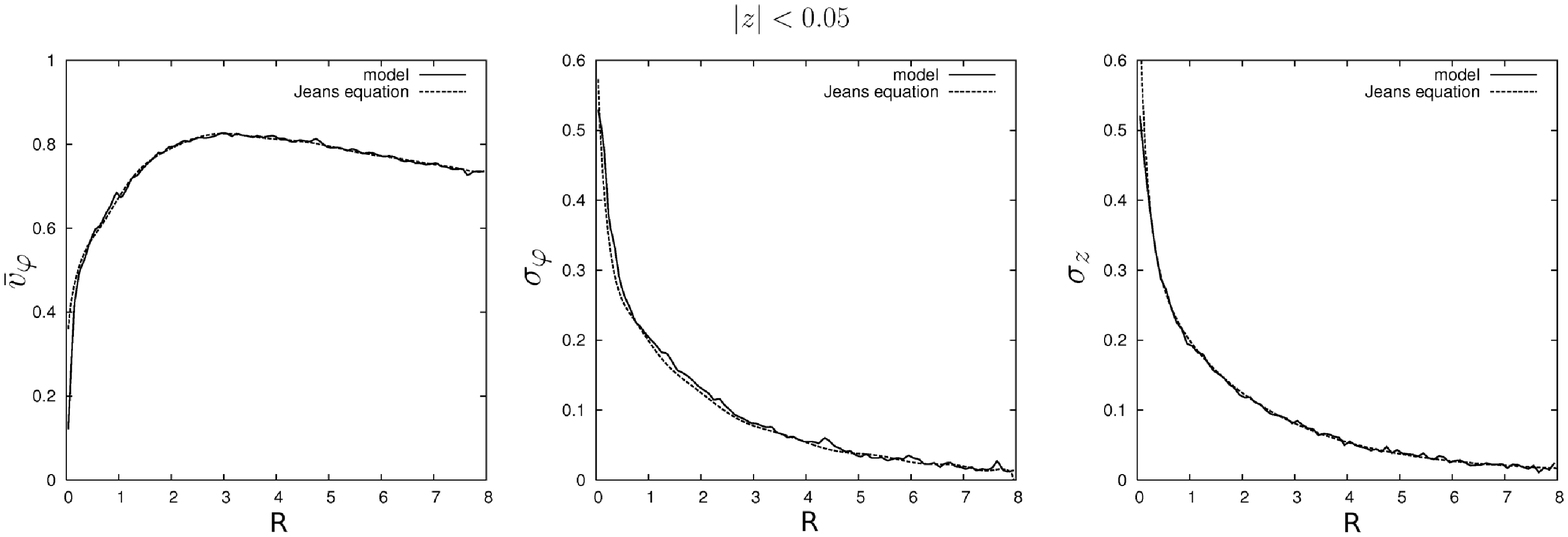}
\end{center}
\caption{
Comparison of profiles of the velocity distribution moments
calculated from the Jeans equations and from the disk of the constructed
disk galaxy model (DISK.SVR). All moments
for the disc were calculated inside the region $|z| < 0.05$.
Left panel:
the solid line corresponds to the value $\bar v_{\varphi}$ for the model,
and the dashed line
corresponds to the same value calculated from the Jeans equation (the
first equation of the system (\ref{eq_jeans})) for $z=0$,
and where the values $\sigma_R$ and $\sigma_{\varphi}$
were taken from
the model. Middle panel: the solid line corresponds to the value
$\sigma_{\varphi}$ for the
model, and the dashed line corresponds to the same value calculated from the
Jeans equation (the second equation of the system (\ref{eq_jeans})),
where the values $\bar v_{\varphi}$ and $\sigma_R$ were taken from the
model.
Right panel: the solid line corresponds to
the value $\sigma_z$ for the model,
and the dashed line corresponds to the same
value calculated from the Jeans equation (the third equation of the system
(\ref{eq_jeans})).}
\label{fig_check_jeans}
\end{figure*}

Fig.~\ref{fig_check_jeans} comparises the radial profiles of
$\bar v_{\varphi}$, $\sigma_{\varphi}$, and $\sigma_z$ calculated from the
constructed disk and from the Jeans equations (\ref{eq_jeans}).
It can be seen that the model follows the Jeans equations very well. 
Note that the moments of velocity distribution both in the Jeans equations 
and in the constructed disk depend on $z$. In
fig.~\ref{fig_check_jeans}, all moments calculated by means of
Jeans equations were calculated for $z=0$. 
We thus took only particles with $|z| < 0.05$ to
calculate them from simulations.
We also checked that our disk follows the Jeans equations very well 
in the rest part of the space (not shown here).

We want to underline that, according the Jeans equations,
the $\sigma_z(R,z)$ in our disk is unambiguously
defined by the chosen mass model of the galaxy (see third equation of system
(\ref{eq_jeans})), as already discussed by \citet{RS06}.

\subsection{Models with given line-of-sight kinematics}
\label{s_losdisks}

\begin{figure}
\begin{center}
\includegraphics[width=7cm]{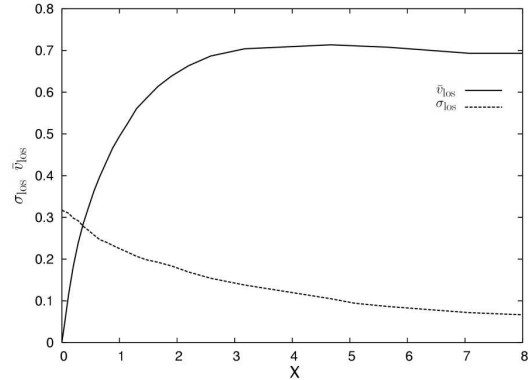}
\end{center}
\caption{The edge-on line-of-sight mean velocity $\bar v_{\rm los}(x)$ and
and velocity dispersion $\sigma_{\rm los}(x)$ for
model DISK.SVR. These parameters are defined in section~\ref{s_fixvlos}.}
\label{fig_losinput}
\end{figure}

 In this section we demonstrate the capability of the iterative method to
construct models with given line-of-sight kinematics. Let us first
calculate the edge-on line-of-sight mean velocity $\bar v_{\rm los}(x)$ 
and the edge-on line-of-sight velocity dispersion $\sigma_{\rm
  los}(x)$ of the 
disk model constructed in the previous section (model DISK.SVR in
section~\ref{s_galaxy}). These profiles are presented on
figure~\ref{fig_losinput}.

We construct two disk models. The first one, called DISK.MVLOS, with a
given $\bar v_{\rm los}(x)$ and the second one, called DISK.SVLOS.
with a given $\sigma_{\rm los}(x)$. Since 
we use the disk galaxy mass model described in the previous section,
the process is similar to reconstructing DISK.SVR by using line-of-sight
kinematic profiles obtained from ``observation''. 

Our initial model for the iterative method was a ``cold'' disk where all
particles move on circular orbits (see previous section where we constructed
DISK.SVR). We made $100$ iterations, each with $t_i=20$. Note that
this is twice the number of iterations used for DISK.SVR, because the
converge in the case of line-of-sight kinematics is slower. The remaining
parameters were taken $dt=1 / 2^4$, $\epsilon=0.04$ and $\theta_t=0.9$. In
order to fix the profile of $\bar v_{\rm los}(x)$ (for DISK.MVLOS)
and the profile of $\sigma_{\rm los}(x)$ (for DISK.SVLOS)
we used the algorithms described in section~\ref{s_fixvlos}.
The number of layers in these algorithms was $n_{\rm div}=200$. 

\begin{figure*}
\begin{center}
\includegraphics[width=14cm]{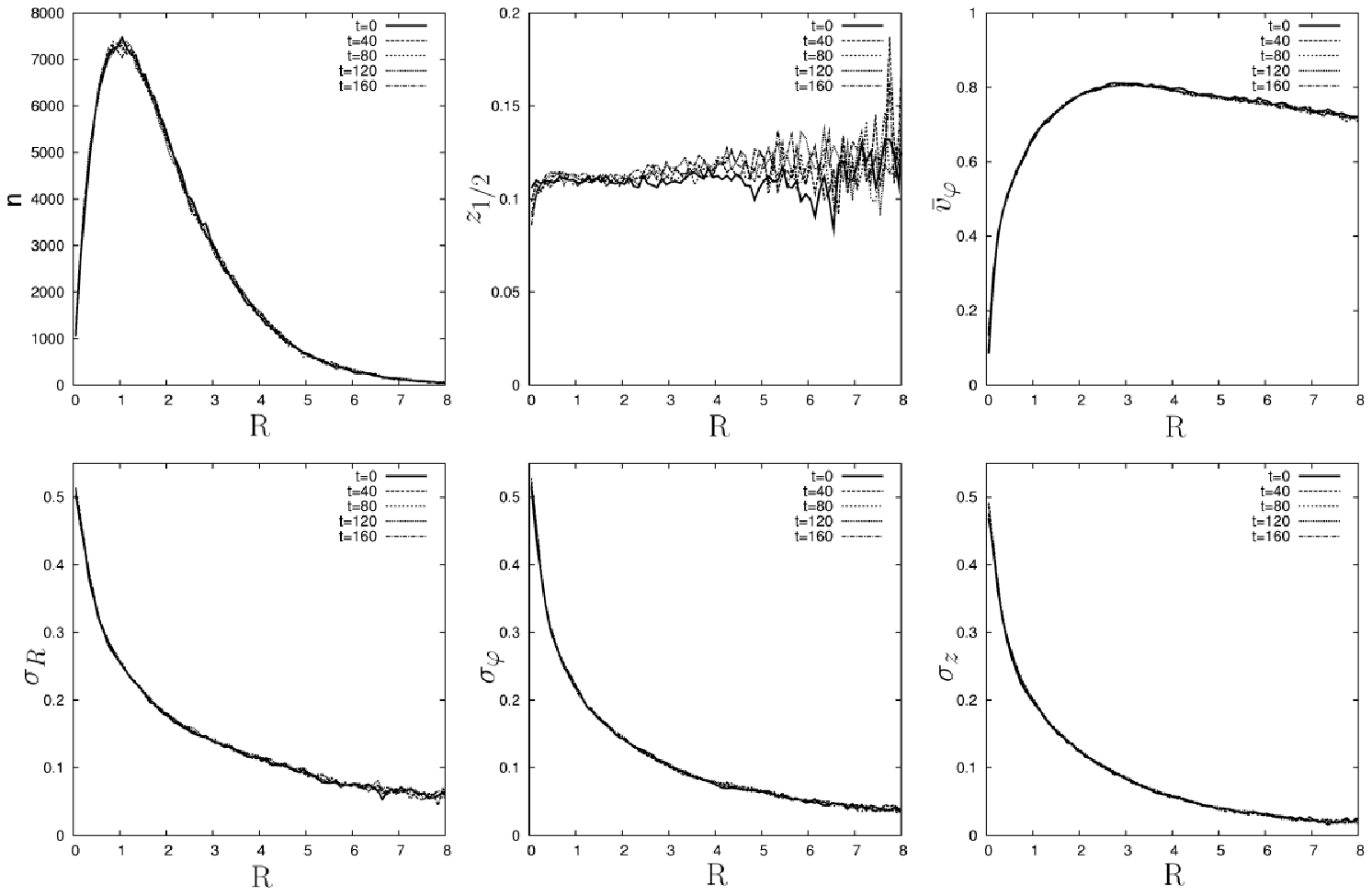}
\end{center}
\caption{Initial evolutionary stages for the model DISK.MVLOS. 
The evolution of the model was calculated with live
disk, halo, and bulge components (we used the halo and bulge constructed in
section~\ref{s_galaxy}).  The same values are
shown as in middle and bottom panels of Fig.~\ref{fig_disk}.}
\label{fig_mvlos_disk}
\end{figure*}

\begin{figure*}
\begin{center}
\includegraphics[width=14cm]{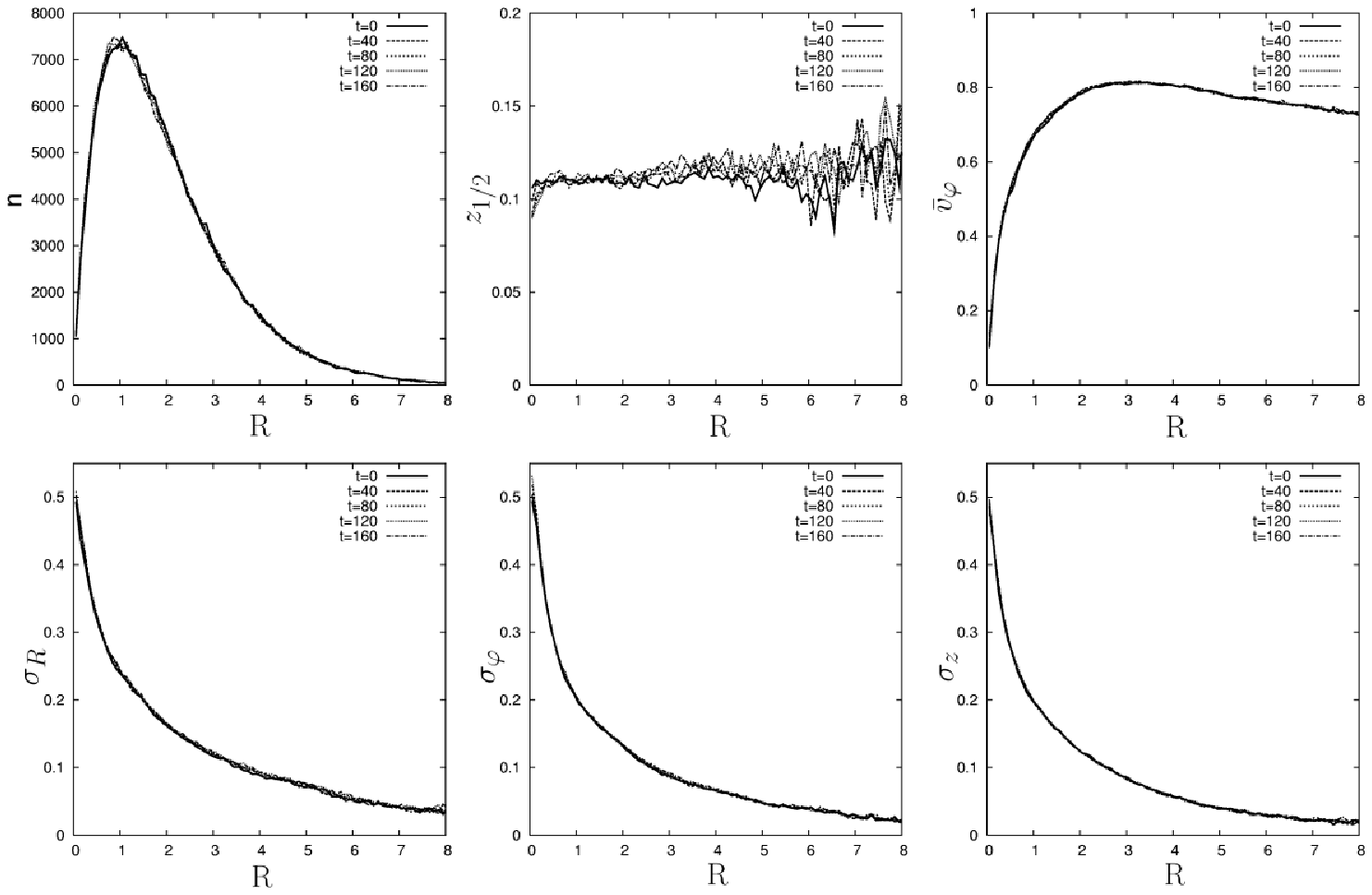}
\end{center}
\caption{Initial evolutionary stages for the model DISK.SVLOS. 
The evolution of the model was calculated with live
disk, halo, and bulge components (we used halo and bulge constructed in
section~\ref{s_galaxy}).  The same values are
shown as in middle and bottom panels of Fig.~\ref{fig_disk}.}
\label{fig_svlos_disk}
\end{figure*}

Let us check the equilibrium of the DISK.MVLOS and the DISK.SVLOS
disks. In both cases we use the halo and bulge
constructed in the previous section, because the mass model is the
same. For the self-consistent evolution we used the same parameters as in previous section. The evolution of the disks in these models is illustrated in 
figures~\ref{fig_mvlos_disk} and~\ref{fig_svlos_disk}, respectively. 
These show that the constructed disks are as close to equilibrium as
they should. 

\begin{figure*}
\begin{center}
\includegraphics[width=14cm]{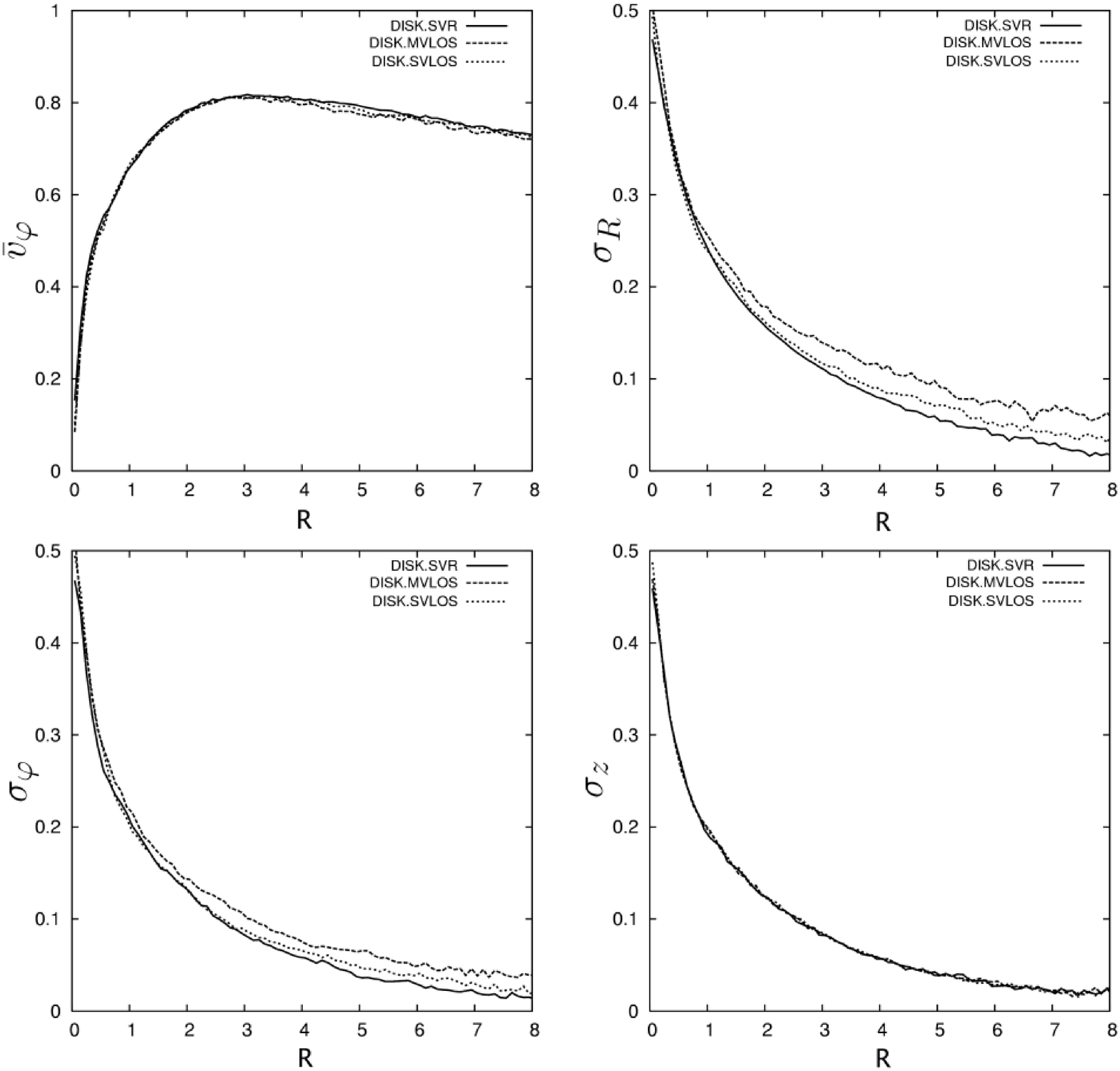}
\end{center}
\caption{The dependence of four moments of the velocity distribution,namely
$\bar v_{\varphi}$, $\sigma_R$, $\sigma_{\varphi}$ and $\sigma_z$
as a function of the cylindrical radius $R$ for DISK.SVR, DISK.MVLOS
and DISK.SVLOS.} 
\label{fig_moments_svR_los}
\end{figure*}

It is interesting to compare DISK.SVR and the two disks constructed
from it by using
line-of-sight kinematics. The three radial profiles of $\sigma_R$
and $\sigma_{\varphi}$ are visibly different, as can be seen in
figure~\ref{fig_moments_svR_los}. The velocity dispersion in the disk
plane is visibly bigger for DISK.MVLOS than for DISK.SVR, especially
near the disk periphery. This is also the case for DISK.SVLOS, but to a
lesser extent. In general, it is clear that both models DISK.MVLOS and
DISK.SVLOS are different from DISK.SVR. This was not expected and must
be due to the fact that more than one equilibrium solution exists for
the adopted constraints. This must be kept in mind when we 
apply our method for constructing phase models of real galaxies and we
will discuss it more extensively in a forthcoming paper. 

\section{Conclusions}
\label{s_conc}
 We presented a new method for constructing equilibrium phase models for
stellar systems --- the iterative method. 
The aim of this method is to construct equilibrium $N$-body models
with given parameters, or constraints. More specifically, these are a 
given mass distribution and, if desired, given kinematic properties, 
parameters, or constraints. Our method is straightforward both
conceptually and in its implementation. We believe that it is this simplicity 
that makes this method so powerful. It simply relies on a
constrained, or guided evolution. We let the
system reach equilibrium via a dynamical evolution in a number of
successive steps. In between two such steps we make sure that the
parameters are set to their desired value and/or that the constraints
are fulfilled. This means that the evolution is guided towards an
equilibrium with the desired parameters and/or constraints. Setting a
mass distribution is of course obligatory, but kinematical constraints
are not. If we wish to include them, we have the choice of a large
number of possibilities, such as setting the radial profile(s) of one, or more
moments of the velocity distribution. 
In this article we described only a few types of kinematic constraints: the
profile of radial velocity dispersion, the profile of velocity anisotropy, a
condition of velocity isotropy and line-of-sight kinematics. 
Procedures for further types of 
kinematic constraints can be easily found following similar techniques.
Furthermore, our implementation of the iterative method can be
directly applied to systems with arbitrary geometry, i.e. the given mass
distribution can be arbitrary and need not have any symmetries. Thus
our method can be used in many different applications.

We used our iterative method to construct several models.
The first one is a triaxial system. The second one is a
multi-component model of a disk galaxy consisting of a stellar disk
with a given
radial velocity dispersion profile, a non-spherical bulge and a halo with
a given anisotropy profile. We also constructed two disk models 
with given line-of-sight kinematic. Using self-consistent
$N$-body simulations,  
we made sure that the models we constructed are indeed very close to
equilibrium (see figs.~\ref{fig_3d},~\ref{fig_disk},~\ref{fig_bulge},
~\ref{fig_halo},~\ref{fig_svlos_disk} and~\ref{fig_mvlos_disk}).

 The iterative method has a number of further applications. It can of
 course be used for constructing equilibrium initial conditions
for $N$-body modelling of stellar systems. For instance, the iterative method
allows one to investigate bar formation in galaxies with a halo having
different kinematics.
Also the iterative method can be applied for constructing phase models of
real galaxies. For example,
we can model observational data by
constructing phase models with given line-of-sight kinematics, as shown 
in section~\ref{s_losdisks}.
This paves the way for studies of e.g. the 
distribution of dark matter in ellipticals, or obtaining phase space
models of observed disk galaxies.
A further interesting application is the study of the properties of 
several equilibrium models for a given mass distribution, as for example
triaxial systems.

The software necessary for the implementation of this method should be
thought of in a very modular way, e.g. with different units for the various
kinematical constraints, and is very straightforward to
write. Nevertheless, we will make our own software publicly available as
soon as this paper is accepted, at the address
http://www.astro.spbu.ru/staff/seger/soft/. This package will
contain also step-by-step examples for constructing models by using 
the iterative method, including the models described in this article. 
Our software uses the $N$-body code gyrfalcON \citep{D00, D02} and the
NEMO package (http://astro.udm.edu/nemo; \citealt{T95}).

\section*{Acknowledgements}
We thank the referee, J. Dubinski, for helpful comments.

This work was partially supported by grants ANR-06-BLAN-0172 and
ANR-XX-BLAN-XXXX, 
by the Russian Foundation for
Basic Research (grants 06--02--16459 and 08--02--00361a)
and by a grant from the President of the Russian
Federation for support of Leading Scientific Schools (grant NSh-8542.2006.02).

\label{lastpage}
\end{document}